\documentclass{Interspeech}

 \interspeechcameraready

\title{Natural Language-based Assessment of L2 Oral Proficiency using LLMs}

\author[affiliation={1}]{Stefano}{Bannò}
\author[affiliation={1}]{Rao}{Ma}
\author[affiliation={1}]{Mengjie}{Qian}
\author[affiliation={1}]{Siyuan}{Tang}
\author[affiliation={1}]{Kate}{Knill}
\author[affiliation={1}]{Mark}{Gales}

\affiliation{ALTA Institute, Machine Intelligence Lab}{Department of Engineering, Cambridge University}{UK}

\email{{\{sb2549,rm2114,mq227,st941,kmk1001,mjfg100\}@cam.ac.uk}}

\usepackage{comment, subcaption, footmisc, multicol, multirow}

\begin{document}

\maketitle

\renewcommand{\thefootnote}{}
\footnotetext{This paper reports on research supported by Cambridge University Press \& Assessment, a department of The Chancellor, Masters, and Scholars of the University of Cambridge.}
\renewcommand{\thefootnote}{\arabic{footnote}}

\begin{abstract}
    
Natural language-based assessment (NLA) is an approach to second language assessment that uses instructions -- expressed in the form of \emph{can-do} descriptors -- originally intended for human examiners, aiming to determine whether large language models (LLMs) can interpret and apply them in ways comparable to human assessment. In this work, we explore the use of such descriptors with an open-source LLM, Qwen 2.5 72B, to assess responses from the publicly available S\&I Corpus in a zero-shot setting. Our results show that this approach -- relying solely on textual information -- achieves competitive performance: while it does not outperform state-of-the-art speech LLMs fine-tuned for the task, it surpasses a BERT-based model trained specifically for this purpose. NLA proves particularly effective in 
mismatched task settings, is generalisable to other data types and languages, and offers greater interpretability, as it is grounded in clearly explainable, widely applicable language descriptors.
\end{abstract}

\section{Introduction}
\label{sec:intro}

The automatic evaluation of language proficiency is gaining increasing attention and significance in the domain of second language (L2) assessment, driven by the steady global rise in the number of L2 learners of English~\cite{howson2013}. Language proficiency exams typically include dedicated sections for assessing speaking, listening, writing, and reading skills, which are traditionally scored by human raters based on specific scoring criteria. Using automated systems for spoken language assessment could provide substantial benefits for teachers, testers, and learners, in formal testing environments as well as in practice contexts in Computer-Assisted Language Learning (CALL). Compared to human raters, automated systems can offer improved consistency and faster feedback at a lower cost, especially considering the high expense and limited scalability associated with recruiting and training human raters~\cite{zhang2013, isaacs2017fully}.

Although not without limitations~\cite{weir2005}, widely adopted frameworks like the Common European Framework of Reference for Languages (CEFR)~\cite{cefr2020} are globally recognised as effective tools for assessing the proficiency of L2 speakers. The CEFR provides proficiency levels organised around \emph{can-do} descriptors that define expected language outcomes at each level. These descriptors encompass multiple dimensions of language ability, such as \emph{grammatical accuracy}, \emph{fluency}, \emph{overall phonological control}, etc. Human raters are typically trained to interpret and apply these descriptors when evaluating L2 learners' performances, ultimately assigning a holistic score. In other words, these holistic scores are based on a modular structure comprising several analytic scores, each reflecting distinct aspects of proficiency. While these analytic scores tend to correlate strongly with holistic scores, they offer different perspectives on a learner's language ability~\cite{dejong2012}.

Existing automatic assessment systems that aim to predict holistic scores typically rely on two approaches: combining handcrafted features into predictive models~\cite{muller2009automatically, crossley2013applications, wang2018towards, liu2020dolphin}, or using end-to-end architectures that leverage transcriptions obtained through speech recognition (ASR) systems~\cite{craighead2020, raina2020universal, wang2021} or the raw audio signal directly~\cite{banno2022proficiency, banno23_slate, mcknight23_slate}. However, these systems provide only a single holistic score and do not offer insights into the underlying components that contribute to that score, limiting their interpretability and usefulness for targeted feedback. In the context of CALL, explainability is particularly critical, as going beyond holistic assessment is supposed to provide much more detailed feedback to L2 learners, by highlighting their fortes and their weaknesses~\cite{hamplyons1995rating}. However, language testers and practitioners are aware that analytic scoring is challenging because raters often struggle to differentiate between specific performance aspects, leading to redundant or biased scores~\cite{lee2009towards}. Furthermore, the high cognitive load involved~\cite{cai2015weight} and the vague definition of some analytic criteria~\cite{douglas1997theoretical} make the assessment process even more difficult for human evaluators and, consequently, hinder the training of automatic systems.

Another persistent challenge in the field of L2 language assessment is the limited availability of publicly accessible high-quality datasets specifically designed for automatic assessment, particularly for spoken language~\cite{knill2024speakimprovecorpus}. This challenge is especially evident with spontaneous speech, which is significantly more difficult to collect and annotate than read-aloud responses.

In light of the current state of automatic speaking proficiency assessment and its associated challenges, we propose natural language-based assessment, an approach to L2 evaluation that leverages natural language proficiency descriptors, originally intended for human raters, in combination with large language models (LLMs) in a zero-shot setting. This approach is particularly well-suited for L2 assessment because LLMs can interpret and apply textual \emph{can-do} descriptors directly to learner-produced data, such as transcriptions, thereby mimicking the interpretative process typically performed by human raters. In doing so, it should offer a way to reconstruct holistic scores through an analysis of underlying performance components, enhancing interpretability and alignment with human assessment principles. Furthermore, by adopting a zero-shot paradigm, this method mitigates the data scarcity problem by capitalising on the broad generalisation and emergent capabilities of LLMs~\cite{wei2022emergentabilitieslargelanguage}. For this purpose, we use an open-source LLM, Qwen 2.5 72B~\cite{qwen2.5}, which processes learner speech transcriptions in conjunction with CEFR analytic descriptors~\cite[pp. 129-142]{cefr2020}. We compare this zero-shot setup to two baselines: a) a BERT-based model~\cite{devlin2018} fine-tuned specifically for this assessment task, and b) Qwen2Audio~\cite{chu2024qwen2audio_etal}, a state-of-the-art speech LLM. The latter is evaluated in two conditions: one fine-tuned on spontaneous spoken data, representing an upper-bound benchmark in an ideal high-resource scenario, and another fine-tuned on read-aloud data to simulate a 
mismatched task
setting, where only easily collected data and annotated data are available. Our results show that the proposed natural language-based assessment method, relying solely on textual input, achieves competitive performance. While it does not outperform Qwen2Audio fine-tuned on spontaneous data, it surpasses the fine-tuned BERT-based baseline and matches the performance of Qwen2Audio trained on read-aloud data.


\section{Natural language-based assessment}
\label{sec:nla}

\begin{figure}[htbp]
    \centering
    \begin{subfigure}[b]{1\linewidth}
        \centering
        \includegraphics[trim=0 16mm 0 16mm, width=1\linewidth]{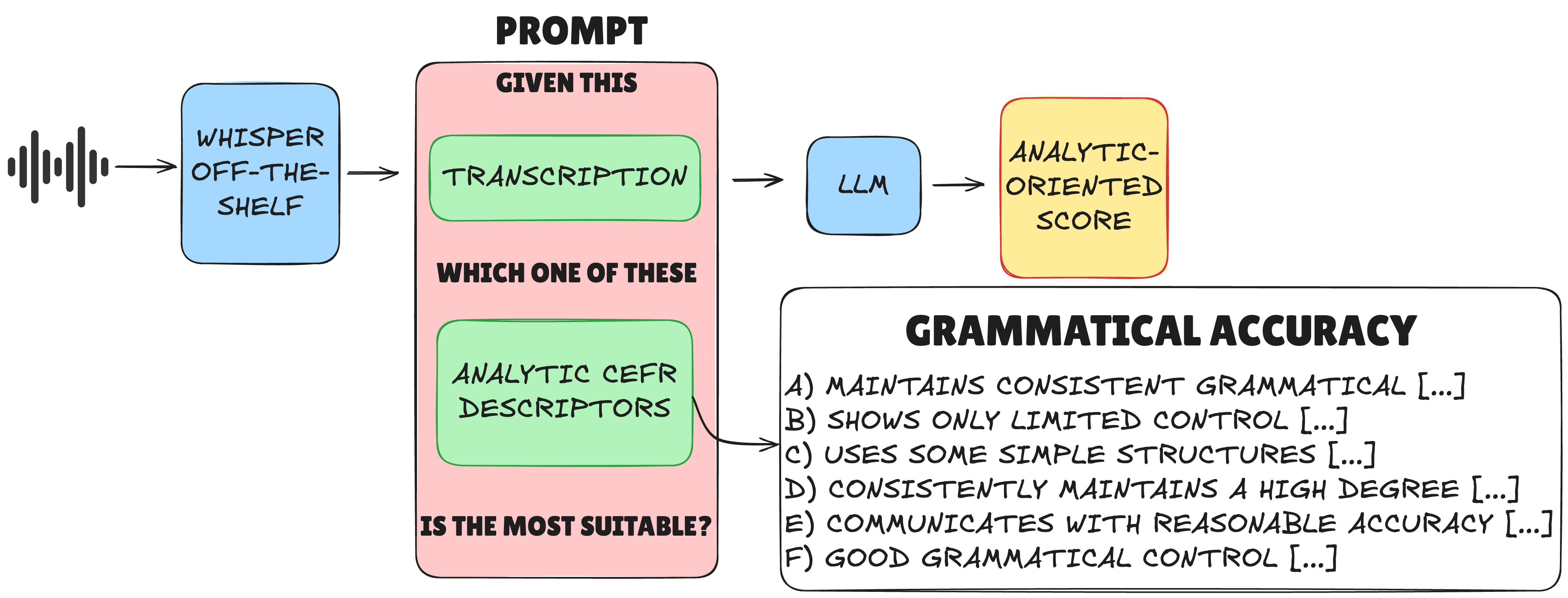}
        \caption{Proposed pipeline (part 1)}
        \label{fig:nla_example_1}
    \end{subfigure}
    \vspace{-6mm}
    \vskip\baselineskip
    \begin{subfigure}[b]{0.9\linewidth}
        \centering
        \includegraphics[trim=0 16mm 0 16mm, width=\linewidth]{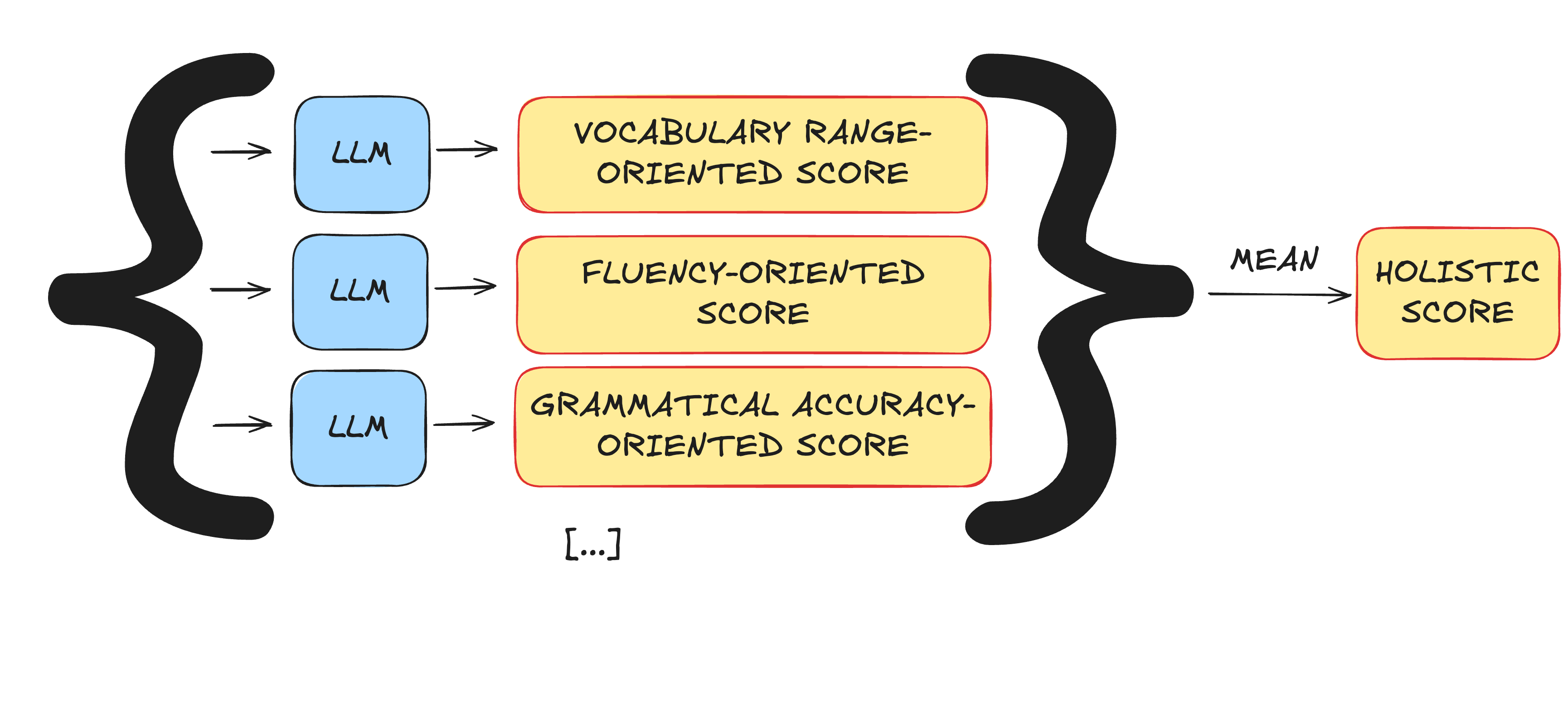}
        \vspace{-10mm}
        \caption{Proposed pipeline (part 2)}
        \label{fig:nla_example_2}
    \end{subfigure}
    \caption{Proposed NLA pipeline for speaking assessment.}
    \label{fig:nla_pipeline}
\end{figure}

Natural language-based assessment (NLA) is an approach to L2 assessment that uses instructions, formulated as scoring criteria or \emph{can-do} descriptors, originally designed for human examiners. It aims to assess whether LLMs can interpret and apply these descriptors in a manner comparable to human raters. We aim to have the LLM replicate the interpretative process typically performed by human raters by generating an analytic-oriented score for each individual proficiency aspect (see Figure \ref{fig:nla_example_1}). These scores are then averaged to reconstruct a holistic score, simulating the aggregation process used in human assessment (see Figure \ref{fig:nla_example_2}). Further details are provided in Section \ref{sec:exp_setup}.

\vspace{-2mm}
A related approach was explored in the context of L2 writing assessment by \cite{yancey-etal-2023-rating}, where the authors examined the use of GPT-3.5 and GPT-4 with holistic CEFR-based descriptors. The models were prompted to select the most appropriate CEFR-based descriptor on a scale from A1 to C2, and the generated responses were converted into numerical scores on a scale from 0 to 6. However, their approach did not involve the use of analytic descriptors. In our previous study on L2 writing assessment~\cite{banno-etal-2024-gpt}, CEFR analytic descriptors were employed not to reconstruct holistic proficiency, but to directly assess specific dimensions of learner performance. In that work, GPT-4 was prompted with analytic descriptors, the learner essay, and the corresponding holistic score, with the goal of producing fine-grained analytic feedback grounded in both the descriptors and the holistic rating. Both works extracted scores by parsing the model-generated responses, rather than employing more sophisticated techniques such as analysing logit probabilities (see Section \ref{sec:exp_setup}). In the domain of speaking assessment, to the best of our knowledge, our recent study~\cite{ma2025slallm} was the first to explore the use of speech LLMs within an NLA framework. Specifically, in that work, Qwen2Audio was fine-tuned to map audio recordings of L2 learner responses to holistic CEFR descriptors. The same model serves as the Qwen2Audio baseline in the present study (see Section~\ref{sec:exp_setup}). It is worth noting that this type of approach requires substantial computational resources. In contrast, the NLA approach proposed in this work leverages speech transcriptions and operates entirely in a zero-shot setting.\footnote{Our previous work~\cite{ma2025slallm} also evaluated Qwen2Audio in a zero-shot setting, though with limited success.}

\section{Data}
\label{sec:data}

This study utilises two datasets: a private dataset, Linguaskill (LNG)~\cite{ludlow2020official}, and a public dataset, the Speak \& Improve Corpus 2025 (S\&I) \cite{knill2024speakimprovecorpus}. Both the LNG and S\&I language tests comprise five sections, each designed to assess a different aspect of a candidate's language proficiency. In Part 1, candidates respond to eight questions, only six of which are scored, as the first two are unmarked. The first four responses are limited to 10 seconds each, while the remaining four allow 20 seconds each. Part 2 is a read-aloud task where the candidate reads eight sentences, each taking approximately 10 seconds. Parts 3 and 4 require candidates to speak for one minute: in Part 3, they express their opinion on a given topic, while in Part 4, they describe a process based on a diagram. In Part 5, candidates answer five topic-related questions, each with a 20-second response time.

\begin{table}[!htbp]
    \centering
    \footnotesize
    \caption{Number of submissions and hours of speech data for each corpus.}
    \begin{tabular}{c|cc|ccc}
    \toprule
    Corpus & \multicolumn{2}{c|}{LNG (P2)} &  \multicolumn{3}{c}{S\&I (P1, 3, 4, 5)} \\
    Split & Train & Dev & Train & Dev & Eval \\
    \midrule
    \#Sub. & 31,476 & 1,033 & 6,640 & 438 & 300 \\
    Hours & 698  & 23  & 244 & 35 & 23 \\
    \bottomrule
    \end{tabular}
    \label{tab:datasets}
\end{table}

Within the S\&I dataset, each section is scored on a scale from 2.0 to 6.0, aligning roughly with CEFR proficiency levels A2 to C. The overall score is calculated as the average of all five sections. For S\&I, a training, development (dev), and evaluation (eval) sets were released as part of the S\&I Challenge 2025 \cite{qian2024speakimprovechallenge}, but they only include Parts 1, 3, 4, and 5. In contrast, for LNG, we focus solely on Part 2 to simulate a 
mismatched task scenario in which only read-aloud data, much easier to collect and annotate than spontaneous data, are available for training. In this case, scores range from 1.0 to 6.0 (i.e., CEFR proficiency levels from A1 to C). Table \ref{tab:datasets} shows the detailed number of test submissions and hours for each corpus. The evaluation is conducted exclusively on the S\&I dev and eval sets, whereas the LNG development set is used solely to optimise one of the Qwen2Audio-based graders (see Section~\ref{sec:exp_setup}).

\section{Experimental setup}
\label{sec:exp_setup}

\noindent \textbf{Qwen 2.5 72B + NLA (Q+NLA):} Our proposed NLA approach utilises an open-source LLM, Qwen 2.5 72B (4-bit quantised)~\cite{qwen2.5}, in a zero-shot setting. As shown in Figure \ref{fig:nla_pipeline}, ASR transcriptions of learner responses generated with OpenAI Whisper small model~\cite{radford2023robust}\footnote{Further details about the word error rate can be found in \cite{qian2024speakimprovechallenge}.} are fed into the model alongside CEFR analytic descriptors for ten individual proficiency aspects, evaluated one at a time: \emph{general linguistic range} (glr), \emph{vocabulary range} (vr), \emph{vocabulary control} (vc), \emph{grammatical accuracy} (ga), \emph{sociolinguistic appropriateness} (soc), \emph{flexibility} (fl), \emph{thematic development} (td), \emph{coherence and cohesion} (cc), \emph{fluency} (flu), and \emph{propositional precision} (pp). Each aspect is rated on a scale from A1 to C2.\footnote{A2 to C2 for \emph{vocabulary control}, \emph{flexibility}, and \emph{thematic development}, as no A1 descriptors are defined for these aspects.\label{footnote_shorter}} While \emph{fluency} can be partially inferred from the presence of disfluencies in speech transcriptions, other acoustically grounded aspects, such as the ones related to \emph{phonological control}, are excluded from our evaluation, as they are not accessible through textual input alone. 
The model is prompted to select the most appropriate analytic descriptor for a given learner response. For each proficiency aspect, the full set of CEFR descriptors is presented using three different random orders, each fixed once chosen, in order to mitigate positional bias~\cite{liusie-etal-2024-llm}. Each descriptor is assigned a corresponding option label (A to F, or A to E for the three aspects without an A1-level descriptor; see Footnote \ref{footnote_shorter}). In other words, A-E are always in the same order, but the order of the descriptors changes. From these options, we extract the logit probabilities and apply a softmax function. Table~\ref{tab:js_divergence} reports the average Jensen-Shannon divergence (JSD) -- a method of measuring the similarity between two probability distributions based on the Kullback–Leibler divergence -- calculated on the softmax probabilities for each individual aspect considering all the examination parts in the S\&I dev set. The results indicate that positional bias does not significantly affect model behavior, except for \emph{general linguistic range}, \emph{grammatical accuracy}, and \emph{sociolinguistic appropriateness}, which appear to be more sensitive to prompt ordering.

\begin{table}[h!]
\footnotesize
\caption{Average JSD values for different aspects.}
\centering
\begin{tabular}{l|c}
\hline
Aspect & JSD \\
\hline

General linguistic range & 0.31 \\
Vocabulary range & 0.08 \\
Vocabulary control & 0.06 \\
Grammatical accuracy & 0.25 \\
Sociolinguistic appropriateness & 0.42 \\
Flexibility & 0.07 \\
Thematic development & 0.15 \\
Coherence and cohesion & 0.16 \\
Fluency & 0.08 \\
Propositional precision & 0.03 \\

\hline
\end{tabular}
\label{tab:js_divergence}
\end{table}

In holistic scoring, human annotators may assign mid-level scores that lie between adjacent CEFR levels, making it challenging to represent them using discrete class labels. To account for this, we adopt \emph{Fair Average}. Let $\mathbf{p} = [p_1, p_2, \dots, p_K]$ denote the softmax probability distribution over $K$ predicted CEFR levels, and let $\mathbf{v} = [v_1, v_2, \dots, v_K]$ represent the corresponding ordinal values assigned to each level (e.g., A1 = 1, A2 = 2, $\dots$, C2 = 6). The \textit{Fair Average Score} is computed as:

\[
\text{FairAvg} = \sum_{k=1}^{K} p_k \cdot v_k
\]

For each proficiency aspect, we obtain three such scores (one for each randomised descriptor order). The holistic score for each part of the exam is computed by averaging all predicted analytic scores. Finally, the overall score for the complete examination is obtained by averaging across all parts.

Additionally, we employ four Ridge regression models ($\alpha=1$)\footnote{We chose Ridge regression (using \texttt{sklearn}) due to multicollinearity among predictors, which can cause redundancy and unstable coefficient estimates in ordinary least squares regression.\label{note_multicollinearity}}, one for each part of the exam, trained on the predicted analytic scores from the S\&I dev set. The models are then evaluated on the eval set, and the resulting per-part predictions are averaged to obtain the overall score. This approach also enables an analysis of the learned regression coefficients, offering insight into which proficiency aspects contribute most to the predicted holistic scores.

\noindent \textbf{BERT-based grader (BERT):} Since our NLA approach relies solely on textual input, a natural and fair baseline is a BERT-based (\emph{bert-base-uncased})~\cite{devlin2018} grading system, following the architecture described in~\cite{raina2020universal, banno23_slate}. Additional implementation details are available in~\cite{qian2024speakimprovechallenge}. We train four separate graders, each corresponding to a specific part of the exam, using ASR transcriptions generated by the OpenAI Whisper small model~\cite{radford2023robust} from the released S\&I training set. Each grader predicts a holistic score for its respective part, and the final overall score is obtained by averaging these four predictions.

\noindent \textbf{Qwen2Audio-based grader (Q2Audio):} We also compare our approach with a set of Qwen2Audio-based graders (\emph{Qwen2-Audio-7B-Instruct})~\cite{chu2024qwen2audio_etal}, trained on audio recordings from the full S\&I training set across all parts, representing our ideal high-resource scenario. In addition, we evaluate another model trained exclusively on LNG Part 2 (i.e., read-aloud data only), simulating a 
mismatched task setting. Further details on the model architecture, training, loss function, and hyperparameters can be found in \cite{ma2025slallm}. For each part of the exam, scores are predicted and then averaged to produce the overall score.

\vspace{1em}

\noindent Results are evaluated in terms of Pearson's correlation coefficient (PCC) and Spearman's Rank Coefficient (SRC).

\section{Experimental results and analysis}
\label{sec:exp_res}

\vspace{-2mm}

\begin{table}[h!]
\footnotesize
 \caption{Results in terms of PCC and SRC on S\&I dev and eval.}
    \centering
    \begin{tabular}{c|c|c|cc|cc}
        \hline
        \multicolumn{1}{c|}{} & Train & Type & \multicolumn{2}{c}{PCC} & \multicolumn{2}{c}{SRC} \\
        \cline{4-7}
        \multicolumn{1}{c|}{} & & & Dev & Eval & Dev & Eval \\
        \hline
        
        BERT  & S\&I & Spont. & 0.753 & 0.727 & 0.764 & 0.728 \\ 
        Q+NLA & - & - & 0.806 & 0.761 & 0.812 & 0.755 \\ 
        Q2Audio & LNG & Read & 0.740 & 0.750 & 0.733 & 0.766 \\
        Q2Audio & S\&I & Spont. & \textbf{0.833} & \textbf{0.821} & \textbf{0.837} & \textbf{0.824} \\
        \hline
    \end{tabular}
    \label{tab:combined_results}
\end{table}

Table \ref{tab:combined_results} reports the overall results considering all the grading systems described in Section \ref{sec:exp_setup}.
It can be observed that Q+NLA outperforms BERT on both the dev and eval sets. On the dev set, Q+NLA significantly surpasses Q2Audio trained on the LNG read-aloud data, while achieving comparable results on the eval set. We observe that the performance of the text-based systems (i.e., BERT and Q+NLA) is lower on the eval set compared to the dev set. This suggests a higher sensitivity of these models to the specific characteristics of the data being analysed. Understanding the root causes of this discrepancy will be a focus of future work. Q2Audio trained under matched conditions remains the best-performing system, but Q+NLA is only a few points behind on the dev set. This is particularly noteworthy given that Q2Audio in matched conditions represents a strong state-of-the-art baseline, while Q+NLA operates entirely in a zero-shot setting and relies exclusively on textual information, without incorporating any acoustic features.

\begin{table}[ht!]
    \centering
    \footnotesize
    \caption{Overall results in terms of PCC and SRC on S\&I eval.}
    \begin{tabular}{c|c|c|c|c}
        \hline
        \multicolumn{1}{c|}{} & Train & Type & \multicolumn{1}{c|}{PCC} & \multicolumn{1}{c}{SRC} \\
        
        \hline
        BERT & S\&I & Spont. & 0.727 &  0.728 \\ 
        Q+NLA & - & - & 0.761  & 0.755 \\ 
        Q+NLA+RR & - & - & 0.771  & 0.760 \\ 
        Q2Audio & LNG & Read & 0.750  & 0.766  \\
        Q2Audio & S\&I & Spont. & 0.821  & 0.824   \\
        
        \hline
        
    \end{tabular}
    
    \label{tab:combined_results_rr}
\end{table}

Additionally, as described in Section \ref{sec:exp_setup}, we train four Ridge Regression (Q+NLA+RR) models on the predicted analytic scores from the dev set and evaluate them on the eval set. This approach yields improvements of roughly 1 point in PCC and 0.5 points in SRC compared to Q+NLA. Beyond performance gains, insightful analysis can be drawn from examining the beta coefficients across the analytic predictions and exam parts, as presented in Table \ref{tab:beta_coeff}.

\begin{table}[ht!]
\footnotesize
\caption{$\beta$ coefficients of each feature across parts. The highest three values in each row are highlighted in bold.}
\centering
\begin{tabular}{l|cccc}
\hline
Analytic score & P1 & P3 & P4 & P5 \\
\hline
$\beta_{\text{glr}}$  & \textbf{0.63} & -0.12 & -0.59 & 0.31 \\
$\beta_{\text{vr}}$   & 0.24 & 0.14 & 0.16 & 0.15 \\
$\beta_{\text{vc}}$   & 0.19 & 0.02 & 0.28 & 0.06 \\
$\beta_{\text{ga}}$   & -0.03 & \textbf{0.63} & \textbf{0.68} & \textbf{0.66} \\
$\beta_{\text{soc}}$  & 0.33 & \textbf{0.98} & 0.25 & -0.49 \\
$\beta_{\text{fl}}$   & 0.05 & 0.42 & -0.06 & 0.05 \\
$\beta_{\text{td}}$   & 0.60 & \textbf{0.51} & 0.37 & \textbf{0.75} \\
$\beta_{\text{cc}}$   & \textbf{0.63} & 0.39 & \textbf{0.71} & \textbf{1.02} \\
$\beta_{\text{flu}}$  & \textbf{1.12} & 0.07 & \textbf{0.94} & 0.60 \\
$\beta_{\text{pp}}$   & -0.59 & -0.22 & -0.22 & -0.06 \\

\hline
\end{tabular}
\label{tab:beta_coeff}
\end{table}

The results indicate that different parts of the exam emphasise distinct communicative competences. In Part 1, which involves short responses to personal questions, \emph{grammatical accuracy} plays a less central role, while \emph{fluency} emerges as the most influential factor. In contrast, Part 3, a long-turn task requiring candidates to express opinions, places greater importance on \emph{sociolinguistic appropriateness}, as well as \emph{grammatical accuracy} and \emph{thematic development}. These findings reflect the demands of the task, where sustained, structured discourse is necessary to discuss advantages and disadvantages. In Part 4, where candidates describe visual prompts such as diagrams, \emph{sociolinguistic appropriateness} is less critical. Instead, \emph{fluency}, \emph{grammatical accuracy}, and \emph{coherence and cohesion} become more salient, likely due to the need for clear and well-organised descriptions. Part 5 requires relatively short responses but still expects candidates to articulate personal opinions. Here, the highest $\beta$ coefficients are observed for \emph{grammatical accuracy}, \emph{thematic development}, and \emph{coherence and cohesion}, with \emph{fluency} also playing a strong role. Interestingly, vocabulary-related features consistently receive lower weights across all parts. This suggests that in this assessment context, lexical diversity and accuracy contribute less to scoring than broader discourse-level or fluency-related abilities.

\begin{figure}[!htbp]
    \centering
    \includegraphics[width=0.85\linewidth]{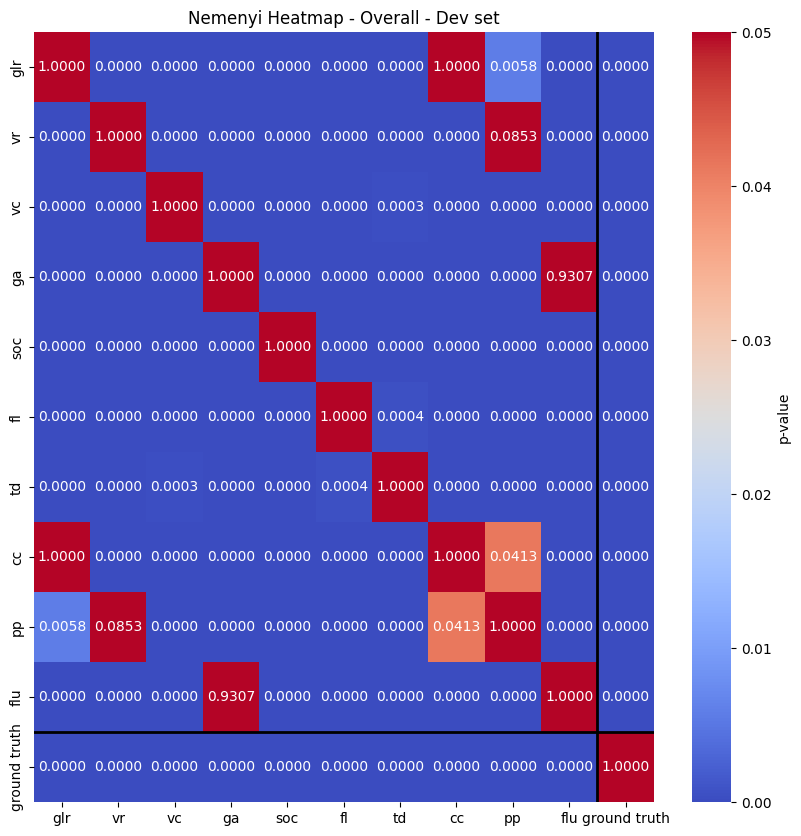}
    \caption{Heatmap of pairwise Nemenyi test between predicted analytic and ground truth scores (overall) on S\&I dev set.}
    \label{fig:nemenyi_test}
\end{figure}

Finally, we examine the relationships among the predicted analytic scores using a repeated measures design. This approach is appropriate because, ideally, we expect differences among the analytic scores, given that they are intended to capture distinct aspects of language performance. Since our data violate the assumption of normality, we adopt the Friedman test~\cite{friedman}, i.e., the non-parametric equivalent of the repeated measures analysis of variance (rANOVA), which assesses whether there are statistically significant differences in the rankings of multiple related groups. The test yields a significant \emph{p}-value, indicating meaningful differences among the predicted analytic scores, despite their underlying intercorrelations (see Footnote \ref{note_multicollinearity}). To identify which pairs of analytic scores differ significantly, we conduct post-hoc pairwise comparisons using the Nemenyi test~\cite{nemenyi}. Figure~\ref{fig:nemenyi_test} presents the results of these comparisons. All paired comparisons, even those with the ground truth score, show significant differences (\emph{p}-value\textless{0.05}), except the pairs \emph{general linguistic range}-\emph{coherence and cohesion}, \emph{vocabulary range}-\emph{propositional precision}, and \emph{grammatical accuracy}-\emph{fluency}. These results suggest that the predicted analytic scores may be capturing some distinct aspects of language proficiency.

\section{Conclusions and future work}
\label{sec:conclusions}

In this paper, we presented our initial experiments on L2 speaking assessment using LLMs in combination with natural language-based analytic descriptors. Our results demonstrate that this approach, relying exclusively on textual information in a zero-shot setting, achieves competitive performance. While it does not outperform speech LLMs fine-tuned for the task, it consistently surpasses a bespoke BERT-based model and matches the performance of a speech LLM fine-tuned on read-aloud data. This highlights the strong potential of our approach, particularly in 
mismatched task settings where no spontaneous training data are available. We have recently applied the method to conversational speech, and preliminary results suggest that it generalises effectively to this 
task, in addition to written data, as shown in prior work. Given the broad applicability of CEFR descriptors across multiple languages, we believe our approach is inherently portable and could be adapted to multilingual assessment tasks, an avenue we plan to pursue in future work. Moreover, the method offers enhanced interpretability, as predictions are anchored in well-established, transparent language descriptors. Our initial statistical analyses indicate promising directions for leveraging these descriptors to better understand the underlying components of holistic scores.

\bibliographystyle{IEEEtran}
\bibliography{mybib}

\end{document}